\documentstyle[12pt,array]{article}
\evensidemargin \oddsidemargin
\catcode `\@=11
\def\numberbysection{\@addtoreset{equation}{section}
         \renewcommand{\theequation}{\thesection.\arabic{equation}}}
\def\be{\begin{equation}}
\def\ee{\end{equation}}
\newcommand{\ba}{\begin{eqnarray}}
\newcommand{\ea}{\end {eqnarray}}
\newcommand{\nn}{\nonumber}
\def\ri{\right}
\def\le{\left}

\def\a{\alpha}
\def\b{\beta}
\def\g{\gamma}
\def\G{\Gamma}
\def\d{\delta}
\def\D{\Delta}

\def\e{\eta}
\def\th{\theta}

\def\L{\Lambda}

\def\rh{\rho}

\def\nil{\emptyset}

\def\ra{\rightarrow}
\def\lra{\longrightarrow}

\def\inf{\infty}
\def\G{\Gamma}

\def\ten{\otimes}

\def\pa{\partial}
\numberbysection
\begin{document}
\begin{center}
   {\bf \huge
   CONCENTRATION FOR ONE AND \\TWO SPECIES
   ONE-DIMENSIONAL \\REACTION-DIFFUSION SYSTEMS \\[25mm] {}}

Horatiu Simon
\\[7mm]
{\it Physikalisches Institut \\ Universit\"at Bonn, Nussallee 12,
53115 Bonn, Germany}
\\[2.2cm]
{\bf Abstract}
\end{center}
\renewcommand{\thefootnote}{\arabic{footnote}}
\addtocounter{footnote}{-1}
\vspace*{2mm}
%
\hspace{\parindent}
We look for similarity
transformations which yield mappings between different one-dimensional
reaction-diffusion processes. In this way results obtained for special
systems can be generalized to equivalent reaction-diffusion models.
The coagulation {\mbox($A + A \ra A$)} or the annihilation {\mbox($A + A \ra
 \nil$)} models can be mapped onto systems in which both processes are allowed.
With the help of the coagulation-decoagulation model results
for some death-decoagulation and annihilation-creation systems are given.
We also find a reaction-diffusion system which is equivalent to the two
species annihilation model  {\mbox($A + B \ra \nil$)}.

Besides we present numerical results of Monte Carlo simulations.
An accurate description of the effects of the reaction rates
on the concentration in one-species diffusion-annihilation model
is made. The asymptotic behavior of the concentration in the two
species annihilation system {\mbox($A + B \ra \nil$)} with  symmetric initial
conditions is studied.
\vfil
\noindent
BONN TH 95-07\\
cond-mat/9503128
\\March 1995
\thispagestyle{empty}
\mbox{}
\newpage
\setcounter{page}{1}
\pagestyle{plain}

%
\section {Introduction}
\label {sec:intro}
\hspace{\parindent}Diffusion controlled reactions in one dimensional
systems have attracted much interest in the last years. They are
non-equilibrium statistical systems which exhibit the property of
self-organization.
Mean field rate equations fail to reproduce the dynamics of these
models which are characterized by non-trivial correlations.
Theoretical descriptions of such systems have to take into account
local fluctuations in the particle density.

Exact results have been obtained for some one-dimensional models in the
continuum from diffusion like equations \cite{bA2}
and on a one-dimensional lattice \cite{Lush, Alc, Pap1}. In
the latter case the time evolution of the system is determined by a
master equation \cite{Kadanoff}. It is useful to rewrite it as a
Euclidian Schr\"odinger equation \cite{Alc}.

Interesting experimental measurements have been reported recently
\cite{S8, S9, S10, S11}. With the help of these experiments
the exciton propagation is studied.  The
conclusion drawn from the measurements is that this propagation  takes
place in one dimension along structures like molecular ''wires''
and chains. The luminescence decay at early times is algebraic:
\be
c(t) \propto \frac{1}{t^{\a}}
\label{alg}
\ee
with $\,\a\,=\,1/2\,$.
This observation yields the conjecture that exciton-exciton annihilation
processes take place \cite{S11}. However, it is not clear whether
exciton-exciton coagulation (fusion) reactions happen. In both the
annihilation and coagulation chemical
models the long-time behavior of the concentration obeys (\ref{alg}).
Moreover, we will prove that this asymptotic behavior of the concentration
is common to all  models in which both reactions take place (\ref{eq:ltb}).
We call them annihilation-coagulation systems. The coefficient of the leading
term depends on the on the ratio of the rates of these two processes.

We show in this paper that all exciton reaction rates can be determined
from experimental data. In our study we use a numerical and an analytical
approach:
\paragraph{First} we want to determine
the next to leading term in the long-time expansion of the concentration
in the coagulation model.
Three results are given in the literature concerning this regime.
An approximative equation for the particle concentration derived from
empty interval probabilities is obtained in \cite{S0}.
A approximation scheme applied to an diffusion-type equations is used in
\cite{S1}.  These two approaches lead to almost the same conjecture:
\be
c(t) = \sqrt{\frac{1}{2\pi t}}+\frac{R}{ \pi
t}\frac{\D'}{1-\D'}\;+O(t^{-x})
\label{eq:ctb}
\ee
where $\,R\,=\,1\,$ in \cite{S1} and $\,R\,=\,\pi/4 \,\simeq \, 0.786\,$
in \cite{S0}.
Here $1- \D'$ is the coagulation rate. The time scale in which the
diffusion rate is equal to one is chosen.
The coefficient of the leading term was exactly computed for $\,\D'\,=\,
\nil$ \cite{Privman}. In this case the next to leading term is of order
$t^{3/2}\;$ \cite{Pap1} and the operator which appears in the master
equation is the Hamiltonian of a supersymmetric system \cite{Alc}.

We verify the conjecture (\ref{eq:ctb}) by determining the value
of the coefficient of the next to leading term in the $\D' \neq \nil$ case.
It can be computed by solving the Bethe Ansatz equations for the $XXZ$
chain in an external field \cite{Alc}. However this computation is a
mathematical challenge because one needs to know both the spectra and the
eigenvectors of the Hamiltonian. Here we investigate this problem with the
help of Monte Carlo simulations. The numerical data presented in Section
\ref{sec:mc} strongly support the $R=1$ conjecture.

\paragraph{Secondly,} we want to find all chemical models which can be
described with the help of (\ref{eq:ctb}) and other results concerning the
coagulation model. We do this by looking for similarity transformations
which map the coagulation model onto other reaction-diffusion systems.
In \cite{Privman, Kang, Spouge} it was shown that the one-species annihilation
and coagulation systems are connected. A general proof is given within the
framework of the ''Hamiltonian'' formalism. The operators which appear in the
master equations describing the two models are the Hamiltonians of quantum
chains \cite{Alc} and are equivalent. We say that two reaction-diffusion
systems
are equivalent when the corresponding Hamiltonians can be obtained
one from the other through a similarity transformation \cite{Pap1}.
In this paper we apply this method and look for further
transformation which connect chemical models and lead to linear relations
among their observables. Thus one can extend results concerning one
reaction-diffusion systems to all its equivalent ones.

We prove that the one-species coagulation-annihilation models can
be divided in equivalence classes with respect to one parameter. Each class
contains also a pure coagulation and a pure annihilation model.

\paragraph{}Another reaction-diffusion system which has attracted much
attention
is the two species annihilation model $(\,A+B\,\lra \nil + \nil \,)$.
No exact results are available for this system.
Scaling considerations indicate that, for equal initial site occupation
probabilities of the two species, the concentration decay is
algebraic with a time exponent equal to $1/4$ \cite{Kang2}.
In \cite{T1} it was conjectured that the asymptotic behavior of the
concentration depends on its initial value:
\be
c_A(t)\,=\,c_B(t)\,=\frac{K\,\sqrt{c(0)}}{t^{1/4}}\,+\,O(t^{-x})\;\;\;\;
\mbox{if}\;\;
c_A(0)=c_B(0)=c(0),
\label{conab}
\ee
where $\,K\,=\,(2\pi)^{-3/4}\,\simeq 0.252\,$. This result is also obtained in
\cite{T3} from a renormalization group analysis for spatial dimensions higher
then two.
A general closure scheme for truncating the hierarchies of the joint
density function equations leads to the same formula
but with $\,K\,=\,(32\pi)^{-1/4}\,\simeq 0.316\,$
\cite{T2}. A value of $\,K\,=\,0.28\,$ has been numerically determined
\cite{T1}. We report results of Monte Carlo simulations which allow
us to obtain a better estimate for the constant $K$. It is
in good agreement with the values conjectured in \cite{T1, T3}.

We also look for the  chemical models equivalent to the two species
annihilation reaction-diffusion system.

The paper is structured as follows:

In section \ref{sec:map} a brief review of the ''Hamiltonian'' formalism is
given. A general condition satisfied by the
transformations which map reaction-diffusion system among themselves
is derived.

In section \ref{sec:res} we look for chemical models equivalent to
systems previously studied. We concentrate on systems for which the
particle concentration was computed.

Two models which the procedure is applied to were mentioned
before: the one species coagulation and the two species
annihilation reaction-diffusion systems.
Another model which is exactly solved in some special case is
the coagulation-decoagulation one
\cite{bA2, Pap1}. We find two reaction-diffusion systems
which are equivalent to it: the annihilation-creation and the
death-decoagulation models.
Thus we can give exact solutions for the particle number in a probabilistic
cellular automaton in which decoagulation and death processes are
permitted, but particles do not diffuse.

In the last section new results obtained from Monte Carlo simulations
are presented.
%
%
\section {The Master Equation and the Similarity Transformation}
\label{sec:map}
\hspace{\parindent} We begin with a brief review of the ''Hamiltonian''
description of reaction-diffusion systems.
We will use the same notations
and conventions as in \cite{Pap1, PRS} where a
comprehensive
description of this formalism is given.

First one has to construct the configuration space. Hard core
interactions among particles moving on the one-dimensional lattice are
considered. This also means that multiple occupation of sites is not allowed.
If the system is populated with one species of particles, each site can be
either empty or occupied by one particle. If there are two species of particles
each site can be in one of three states. The number of states is denoted by $N$
and the lattice length by $L$.
To each site $i$ we attach a $N$-dimensional vector
space $V_i$.
In the two state models ($N=2$) the vector $(\,1,\;0\,)$
corresponds to a vacancy while the vector $(\,0,\;1\,)$ corresponds to
a occupation of the site by an $A$ particle. In the three state models
($N=3$)
the vectors $(\,1,\;0,\;0\,)$, $(\,0,\;1,\;0\,)$ and $(\,0,\;0,\;1\,)$
correspond to an empty site and
to occupation of the site by a particle $A$ or $B$ respectively.

The configuration space
has the structure of an $L$-fold tensor product $V = V_1\ten\cdots\ten V_L$.
Each of the $N^L$ vectors of the tensor product base corresponds to a possible
configuration of the lattice. A natural convention used to denote these
vectors is obtained by associating to each site a variable $\b_{i}$ which
takes integer values between $0$ and $N-1$.  In the two state models we use
$\b_{i}=0$ to index a vacancy while $\b_{i}=1$ corresponds to the presence
of an $A$ particle. In addition, $\b_{i}=2$ is used to describe a site that is
occupied by a $B$ particlei, in the three state models.  In this 'spin basis'
\cite{Kadanoff}
\be
|\{\b\}\rangle = |\b_1,\ldots,\b_L\rangle
\label{base}
\ee
we define the ket vector
\be
|P\rangle = \sum_{\{\b\}}P(\{\b\};t)|\{\b\}\rangle
\ee
which describes the state of the system. $P(\{\b\};t)$ stands for the
realization probability of the configuration $\{\b\} =
\{\b_1,\ldots,\b_L\}$ at time $t$.

The state of a
system can also be described by the set of empty intervals probabilities
\cite{Doering62}. They
are defined as the probabilities to find one or more sequences of
consecutive vacant sites. This description sometimes simplifies exact
computations.

The dynamics of reaction-diffusion systems are determined by the
rates of the allowed processes. We consider only reactions which  can take
place when two hopping particles collide. So the set of parameters
is given by the
probabilities that a state $(\a,\b)$ on two adjacent sites will change into
the state $(\g,\d)$ after one unit of time, denoted by

\be
\G^{\a,\b}_{\g,\d}\;\;\;\;\mbox{with the convention}\;\;\;\;
\G^{\a,\b}_{\a,\b}=0.
\ee
Here $\a,\b,\g$ and $\d$ take values between $0$ and $N-1$.
The sum of all reaction rates which modify the state $(\a,\b)$ of
two neighboring sites is denoted by
\be
\G_{\a,\b}=
\sum_{\g,\d =0}^{N-1}\!\!\!\,
                   \G^{\a,\b}_{\g,\d}\;.
\label{nochange}
\ee
All rates are non-negative, constant and real.
We discuss only the left-right symmetric case. That is all the rates satisfy
the condition
\be
\G^{\a,\b}_{\g,\d}\;=\;\G^{\b,\a}_{\d,\g}.
\ee
For one species systems a list of reactions and corresponding
rates is given below:
\ba
\mbox{diffusion} & A + \nil \ra \nil + A \; & \mbox{rate} \;\;\G^{10}_{01}
\nonumber \\
\mbox{annihilation} & A + A \ra \nil + \nil \; & \mbox{rate} \;\;\G^{11}_{00}
 \nonumber \\
\mbox{creation} & \nil + \nil \ra A + A \; & \mbox{rate} \;\;\G^{00}_{11}
\nonumber \\
\mbox{coagulation} & A + A \ra A + \nil \; & \mbox{rate} \;\;\G^{11}_{10}
\nonumber \\
\mbox{decoagulation} & A + \nil \ra A + A \; & \mbox{rate} \;\;\G^{10}_{11}
\nonumber \\
\mbox{death} & A + \nil \ra \nil + \nil \; & \mbox{rate} \;\;\G^{10}_{00}
\nonumber \\
\mbox{birth} & \nil + \nil \ra A + \nil \; & \mbox{rate} \;\;\G^{00}_{10}.
\nonumber
\ea

The master equation \cite{Kadanoff}  describing the time evolution of the
probability distribution $P(\{\b\};t)$ can be written in the form of a
Euclidian Schr\"odinger equation \cite{Alc}
\be
\frac{\pa}{\pa t}|P\rangle = - H|P\rangle\;.
\label{eqn:schro2}
\ee
$H$ operates on $V$. We consider periodic boundaries.
Due to the fact that the reactions take place only between particles
placed on two
neighboring sites this Hamiltonian
can be written as a sum:
\be
H\;=\,\sum_{i=1}^L \, H_i \;,
\label{hsum}
\ee
where $ H_i$ acts locally on $V_i\ten V_{i+1}$. This operator
is defined by:
\ba
H_i  & = & \sum_{\a,\b=0}^{N-1}\;
              \bigl [\G_{\a,\b}\, E^{\a,\a}_{i}E^{\b,\b}_{i+1}
         -\,\sum_{\g,\d=0}^{1}\!\!\!\;
                \G^{\a,\b}_{\g,\d}\, E^{\g,\a}_{i}E^{\d,
\b}_{i+1}\,\bigr ]\;,
\label{eqn:Hi}
\ea
where $E^{kl}$ are $N \times N$ matrices with the entries
$(E^{kl})_{nm}=\d_{k,n}\d_{l,m}$. This Hamiltonian is not necessarily
hermitian. It ensures
the conservation of
probability and thus
has a bra ground state
\be
\label{bra0H0}
\langle 0|H=0
\ee
which is the sum of all basis vectors
\be
\langle 0| = \sum_{\{\b\}}\langle\{\b\}|\,.
\label{eq:brags}
\ee
In other words : if we consider $H_i$ as a matrix, equation (\ref{eqn:Hi})
with (\ref{nochange}) states that the sum of the entries
from each column is $0$. $H$ has the same property (\ref{bra0H0}).

The formal solution of (\ref{eqn:schro2}) is:
\be
|P(t)\rangle = \exp(-Ht)|P_0\rangle\,
\ee

where the initial state is denoted by $|P_0\rangle$ \cite{Alc}.

The expectation value of an observable $X$ is given by the matrix element
\be
<X>(|P_0\rangle, t) = \sum_{\{\b\}}X(\{\b\})P(\{\b\};t)\;=\;\langle 0|X|P(t)
\rangle\;.
\label{eq:exval}
\ee

The advantage of this Hamiltonian formalism is that we can introduce
similarity transformations.  Our interest is to find those $B$ for which the
transformed Hamiltonian
\be
\hat{H}\; =\;\sum_{i=1}^L \, \hat{H}_i\;=\;B\,H\,B^{-1}
\label{newh}
\ee
describes also the dynamic of a reaction-diffusion system.
(The quantities in the transformed model are denoted with a $\;\hat{}\;$.)
Therefore we have to concentrate on local transformations:
\be
B = b\ten b\ten\cdots \ten b
\label{local}
\ee
defined by the $L$-fold tensor product of a real $N\times N$ matrix $b$.
The mapping can easily be extended to the computation of expectation values.
They can be computed \cite{Pap1} from the corresponding expectation values in
the 'original' model:
\ba
\widehat{<X>}(|P_0\rangle,t) =
\langle 0|X\exp(-\hat{H} t)|P_0\rangle =\nonumber \\
\nonumber \\
\langle0|XB \exp(-Ht) B^{-1}|P_0\rangle =
<XB>(B^{-1}|P_0\rangle,t)
\label{eq:exvalnew}
\ea

Some restrictions have to be imposed on $b$. First, the sum of the entries
from each column of $\hat{H}_i$ must be $0$:
\be
\label{bra0BH0}
\langle 0|BH_i=0\;.
\ee
A general solution of (\ref{bra0BH0}) is:
\be
\langle 0|B = \rh \langle 0|
\label{bra0Ba0}
\ee
Here $\rh$ is real. We have checked explicitly for every model considered
in this paper that transformations obeying (\ref{bra0BH0}) but not
(\ref{bra0Ba0}) do not lead to new mappings.
Using (\ref{eq:brags}) and (\ref{local}) one gets the restriction:
\be
\sum_{i=0}^{N-1}b_{ij} = \rh^{\frac{1}{L}}\;.
\label{consu}
\ee
We can eliminate a overall scaling parameter by choosing $\rh =1$.
Thus the transformations are represented by real $N \times N$ matrices with
the property that the sum of the entries on each column is equal to $1$.

Not all these  $N \times (N-1)$ - parameter transformations lead to another
reaction-diffusion system. The matrix elements of the two sites $(\,L=2)\:$
Hamiltonian $\hat{H}$ have to satisfy the following sign conditions:
\ba
\hat{H}_{mm} & \geq 0 & m = 0,1 \ldots N^2-1  \nonumber \\
\hat{H}_{mn} & \leq 0 & m,n = 0,1 \ldots N^2-1,\; m \neq n
\label{sign}
\ea
The remaining transformations map reaction-diffusion system one onto the other.
The program {\em REDUCE} was used to compute the transformed Hamiltonians.

\section {Mappings between reaction-diffusion systems}
\label{sec:res}
\hspace{\parindent} We will start with the Hamiltonians of those
chemical models for which the behavior of the particle concentration
is known and look for local transformations
which map them onto other reaction-diffusion systems.

There are always the $N!\,-\,1\,$ permutation transformations which
interchange (if $N=3$ different types of) particles and vacancies.
We will treat everything ''modulo'' these permutations.

As we mentioned before, we restrict ourselves to the left right symmetric case.
The similarity transformation preserves this property. The
generalization to the non-symmetric case is straightforward.

In the three state model we take equal
diffusion rates for the two kinds of particles
($\G^{01}_{10}\,=\,\G^{02}_{20}$).
One parameter of the problem can be eliminated  by choosing the time scale
in which
the diffusion rates
are equal to 1.

\subsection {Two-state models}
\label{2st}
\hspace{\parindent}
Exact results are available for the one-species annihilation,
coagulation
and coagulation-decoagulation models. For the last two models the system of
linear differential equations for the empty interval probabilities was
written and solved. These equations decouple
from those corresponding to many intervals probabilities \cite{PRS, DBH}.
In the transformed models identical systems of equations
can be written but for other observables, namely the probabilities of
having interpolating sequences of vacancies and particles on consecutive
sites \cite{PRS}.

The general form of the local transformation is given by (\ref{consu}):
\vspace{0.5cm}
\be
b = \le( \begin{array}{cc}
         1-\a & \b\\
         \a   & 1-\b
    \end{array} \ri)\;
\label{genf2}
\ee

We will start with the annihilation model and not with the equivalent
coagulation one. For simplicity we will use the notation $a\;$, instead of
$\G^{11}_{00}$, for the annihilation rate and $f\;$ for the coagulation
rate ($\G^{11}_{10}$).
\paragraph{The annihilation model.} In this case particles diffuse and
disappear
pairwise when two of them try to occupy the same site. We denote the
corresponding Hamiltonian with $H_{\mbox{\scriptsize ann.}}$
and introduce, as in \cite{Alc, Pap1, Pap2}, the parameter $\D'$ defined by:
\be
\D' = 1-\frac{a}{2}\;.
\label{eq:DelA}
\ee
If $\,a\,<\,2,\;\D'\;$ is equal to the probability that no reaction
takes place at an two particles encounter. If the parameter $\D'$ is
zero exact solutions can be obtained because $H_{\mbox{\scriptsize
ann.}}$ can be expressed in terms of free fermions. In \cite{Lush}
the time dependence of the particle concentration is derived in the
thermodynamical limit with a full lattice as initial configuration.
The finite chain is treated exactly in \cite{Alc} and the two point
correlation functions for the infinite chain are obtained. The particle
concentration for random homogeneous initial conditions is deduced in
\cite{Pap1}.

It can easily be seen that the transformed Hamiltonian $\hat{H}$ will only
fulfill the conditions (\ref{sign}) if at least one of the entries of $b$ is
nonzero. We take $\a=0$. The case $\b = 0$ can be obtained by applying a
permutation transformation on $\hat{H}$.

It is a well known fact \cite{Privman, Kang, Spouge} that the
annihilation and coagulation models are equivalent.
We reobtain this result by choosing $\b=-1\;$.
In this case $\hat{H}$ is identical with the Hamiltonian of the
coagulation model \cite{Pap1}.

This is, however, only one of the possible choices for the transformation
parameter. For $-1 \leq \b \leq 0\;$ we get the Hamiltonian of a system in
which apart from diffusion with the same rate $\hat{\G}^{10}_{01}\,=\,1$,
the following processes are permitted:
\begin{itemize}
\item annihilation with rate
\[\hat{\G}^{11}_{00} = \hat{a} = 2\frac{\b+1}{1-\b}(1-\D') \nonumber\]
\item coagulation with rate
\[\hat{\G}^{11}_{01} = \hat{f} = 2\frac{\b}{\b-1}(1-\D')\]
\end{itemize}

The matrix $b$ and $\D'\;$ can be written as functions of $\hat{a}$
and $\hat{f}$:
\be
\D'=1-\hat{f}-\hat{a}/2
\label{d'}
\ee
or equivalently $a=\hat{a}+2\hat{f}$. One gets the following result:
\[
\hat{H}(\hat{a},\hat{f})=BH_{\mbox{\scriptsize ann.}}B^{-1}
\]
where
\be
b = \le( \begin{array}{cc}
         1 & \frac{-\hat{f}}{\hat{a}+\hat{f}}\\
         0 & \frac{\hat{a}+2\hat{f}}{\hat{a}+\hat{f}}
    \end{array} \ri)\;
\label{part2}
\ee
and with the parameter $\D'$ (\ref{d'}).

We have obtained a one-parameter group of similarity transformations
which act between annihilation-coagulation models ($(a,f)\lra (\hat{a},
\hat{f})$) corresponding to the same value of $\D'$ (i.e. $a/2+f=\hat{a}/2
+\hat{f}$ ). It is straightforward to write the transformation which connects
the coagulation model $(a=\nil)$ and one described by
$\hat{H}(\hat{a},\hat{f})$.

Some explicit results concerning the particle concentration
can be given now by applying transformations (\ref{eq:exvalnew})
on formulas  obtained for the annihilation ($f=0$) (\ref{part2})
or coagulation ($a=0$) models.

Uncorrelated initial conditions are considered: at time $t=0$
every site is occupied with  the same probability $p$. In this case:
\[
B^{-1}|\vec{P}_0\rangle =
{{1-\frac{\hat{a}+\hat{f}}{\hat{a}+2\hat{f}}p}\choose{\frac{\hat{a}+\hat{f}}
{\hat{a}+2\hat{f}}p}}_1\ten\cdots\ten
{{1-\frac{\hat{a}+\hat{f}}{\hat{a}+2\hat{f}}p}\choose{\frac{\hat{a}+\hat{f}}
{\hat{a}+2\hat{f}}p}}_L
\]
which corresponds to an initial site occupation probability in the
equivalent annihilation model equal to $\frac{\hat{a}+\hat{f}}{\hat{a}+2\hat{f}
}p$.

The operator of the occupation number of the
site $i$ is given by:
\be
n_i = \le( \begin{array}{cc}
         0 &  0\\
         0 &  1
    \end{array} \ri)_i\;.
\ee
The expectation value of the product of $k$ such operators
in the coagulation-annihilation model is
(\ref{eq:exvalnew}):
\be
\widehat{<n_{i_1}\cdots
n_{i_k}>}(p,t)=\le(\frac{\hat{a}+2\hat{f}}{\hat{a}+\hat{f}}
\ri)^k<n_{i_1}\cdots
n_{i_k}>_{\mbox{\scriptsize ann.}}\le(\frac{\hat{a}+\hat{f}}{\hat{a}+2\hat{f}}
p,t\ri)
\label{cag}
\ee

We now drop the superscript.

The exactly solvable case $\D'\,=\, 1-f-\frac a 2 \,=\,0$ corresponds to the
physical situation that whenever two particle meet at least one of them
disappears from the system.
We mentioned in the introduction that the operators which appears in the master
equation of these models are the Hamiltonians of supersymmetric systems.
This reaction-diffusion models are mentioned in \cite{Derr} in connection
with the Glauber dynamic of the $q$ state Potts model.
The results given in \cite{Pap1} for $f=1$ or $a=2$
can be generalized to these annihilation-coagulation models. The particle
concentration on a finite chain is:
\ba
c(t,p,L) &=&
  \frac{1-[1-(2-f)p]^L}{(2-f)\,L} \nn\\[2mm]
& &  -\;\frac{1}{(2-f)\,L}\;\sum_{k=1}^{L-1}\bigg\{
\biggl[\frac{1-(-1)^{k}[1-(2-f)p]^L}
              {\frac{1+[1-(2-f)p]^{2}}{2[1-(2-f)p]}-\cos{\frac{\pi
k}{L}}}
        -\frac{1-(-1)^{k}[1-(2-f)p]^L}{1-\cos{\frac{\pi
k}{L}}}\biggr]\nn\\[2mm]
& &  \hspace{20mm}\times\sin^{2}{\biggl(\frac{\pi k}{L}\biggr)}
      \exp{\biggl[-4t\biggl(1-\cos{\frac{\pi k}{L}}\biggr)\biggr]}\biggr\}\;.
\label{eq:genkonz}
\ea

The behavior of the concentration (\ref{eq:genkonz}) in the finite-size
scaling limit \cite{Alc}:
\be
L\rightarrow\inf,
\;\;t\rightarrow\inf\;\; \mbox{with}\;\; z=\frac{4t}{L^2}\;\; \mbox{fixed}
\label{fssl}
\ee
can be determined \cite{Pap1}.
Exact expressions can be written for the scaling and the first correction
functions in terms of Jacobi theta functions. We get the following
scaling relation:
\ba
L\:c(z,p,L) & = &\frac{1}{2-f}\bigg\{\th_3\:\biggl(0,
\frac{i\pi z}{2}\biggr)
+\frac{1}{L^2}\:\biggl[\:\frac{z}{6}
                 \:\frac{\partial^2}{\partial z^2}\:
                  \th_3\:\biggl(0,\frac{i\pi z}{2}\biggr) \nn\\
   & &               +\frac 1 2 \biggl(1-\frac{2}{(2-f)p}\biggr)^2\:
			\frac{\partial}{\partial z}\:
                  \th_3\:\biggl(0,\frac{i\pi z}{2}\biggr) \biggr]\:
	\biggr\}\;
              +O\:\biggl(\frac{1}{L^4}\biggr).
\label{eq:scf}
\ea

Another interesting result concerns the large time behavior of the
particle concentration in the thermodynamic limit. In the $\D'\,\neq\,
\nil\,$ case we start from the conjectured expression (\ref{eq:ctb}).
Applying (\ref{eq:exvalnew}) we get:
\be
c(t) = \frac{a+2f}{a+f}\le[\sqrt{\frac{1}{8\pi t}}+\frac{R}{2\pi
}\frac{\D'}{1-\D'}\frac{1}{t}\ri]\;+O(t^{-x})
\label{eq:ltb}
\ee

The coefficient of the leading term depends only on the ratio of
the annihilation and the coagulation rates. The next to
leading term ($\sim \frac 1 t \;$) is dependent of $\D'$. Both of them are
independent on the initial concentration. We will give strong numerical
evidence which supports the conjecture presented in \cite{S1} (i.e.
$R=1$) in Section (\ref{sec:mc}).

This asymptotic expression is an exact result in the supersymmetric ($\D'=0$)
case \cite{Lush}. The next to leading term is of order $t^{-3/2}$.
For uncorrelated homogeneous initial conditions, in the $z \rightarrow
\nil$ limit \cite{Pap1} of (\ref{eq:scf}) one gets:
\ba
c(t,p) &=& \frac{1}{2-f}\bigg\{\sqrt{\frac{1}{2\pi t}}  \nn\\
& &  -\frac{1}{\sqrt{2\pi}}
\frac{1}{16}\le[\biggl(1-\frac{2}{(2-f)p}\biggr)^2\: - \frac 1 2 \ri]\frac {1}
{t^{3/2}}\biggr\}\;+O(t^{-5/2})
\label{eq:slt}
\ea

\paragraph{The coagulation-decoagulation model.}
In this system, apart from diffusion, coagulation and its backward reaction
are permitted.
We will use, as in \cite{Pap1}, the variable $\e$ defined through
the decoagulation rate:
$\e^2\,=\,\G^{10}_{11}+1$.

Exact solutions of this model are available in the continuum limit
\cite{bA2} and for the finite lattice in the case in which the
coagulation and diffusion rates are equal ($f=1$) \cite{Pap1}.
In the latter case the steady state concentration is $c_s=1-\e^{-2}$,
independently of $L$ and $p$.

We found that the coagulation-decoagulation model is equivalent with two
other reaction-diffusion systems. Only two transformations, which correspond
to special choices for $\a$ and $\b$ in (\ref{genf2}), connect this model
with others.

a) One of them is:
\be
b = \le( \begin{array}{cc}
         1 & -\frac f {\e^2-1} \\
         0 & 1+\frac f {\e^2-1}
    \end{array} \ri)\;
\label{codeco}
\ee

If $f \leq 1$, the transformed Hamiltonian $\hat{H}$ corresponds to a
reaction-diffusion system with the following processes:

\begin{itemize}
\item diffusion with rate $\hat{\G}^{10}_{01}\,=\,1-f$
\item decoagulation with rate $\hat{\G}^{10}_{11}\,=\,f+\e^2-1$
\item death
with rate $\hat{\G}^{10}_{00}\,=\,f$
\end{itemize}

{\em If} $f=1$ {\em there is no diffusion in the new model. The motion of the
particles is realized by successive processes of decoagulation (with
rate }$\e^2\,>\,$ {\em 1 ) and death (with rate 1) on neighboring pairs
of lattice sites.}
In this case we can use the results from \cite{Pap1}.
For the finite chain and homogeneous random initial conditions the particle
concentration is:

\ba
c(t,p,L) & = & \frac{1-[1-p(1-\e^{-2})]^L}{1-\e^{-2L}} \nn\\[2mm]
& &          -\;\frac{1}{\e L}\;\sum_{k=1}^{L-1}\bigg\{\biggl[
    \frac{1+(-1)^{k+1}\e^L[1-p(1-\e^{-2})]^L}{\frac{1+\e^2[1-p(1-\e^{-2})]^{2}}
{2\e[1-p(1-\e^{-2})]}-
\cos{\frac{\pi k}{L}}}-\frac{1+(-1)^{k+1}\e^L[1-p(1-\e^{-2})]^L}{\frac{\e^2+1}
{2\e}
-\cos{\frac{\pi k}{L}}}\biggr]\nn\\[2mm]
& & \hspace{2cm} \times \, \sin^{2}{\biggl(\frac{\pi k}{L}\biggr)}\exp
{(-\L^0_kt)}\biggr\}\;.
\label{eq:coagkonz}
\ea
where $\L^0_k$ is:
\be
\L_k^0
  =  \e\biggl(2(\e+\e^{-1})\:-4\:\cos\frac{\pi k}{L}\biggr)\;.
\ee

The full lattice is a stationary state:
\[
c(0,1,L)\,=\,c(t,1,L)\,=1.
\]

Interchanging $A$ and $\nil$ a system with a death rate $\, \e^2 \,$
greater than the decoagulation rate $\, 1 \,$ and zero steady state
concentration is obtained.
For this choice of rates (\ref{eq:coagkonz}) gives
the concentration of vacancies if $p$ is replaced by $1-p$.

\vspace{0.5cm}

b) The other transformation which gives a mapping of the
coagulation-decoagulation model is:

\be
b = \le( \begin{array}{cc}
        \frac 1 2 \biggl[1+\sqrt{ \frac{\e^2-1}{f} +1}\biggr] \;&\; \frac 1
2\\
&\\
        \frac 1 2 \biggl[1-\sqrt{ \frac{\e^2-1}{f} +1}-1\biggr] \;&\; \frac 1
2
    \end{array} \ri)\;
\label{ancret}
\ee

One obtains an annihilation-creation reaction-diffusion system with rates:
\begin{itemize}
\item diffusion $\hat{\G}^{10}_{01}\,=\,\frac{\e^2+1}{2}$
\item annihilation $\hat{\G}^{11}_{00}\,=\,\hat{a}\,=\,\frac 1 2 \biggl[\sqrt{
\frac{\e^2-1}{f} +1}+1\biggr]^2 f^2$
\item creation
$\hat{\G}^{00}_{11}\,=\,\frac 1 2 \biggl[\sqrt{ \frac{\e^2-1}{f}
+1}-1\biggr]^2 f^2$.
\end{itemize}

The annihilation rate is greater than the creation rate.
The time unit can be adjusted in such a way that the new
diffusion rate $\hat{\G}^{10}_{01}$ becomes equal to $1$.

In the case $f=1$ the sum of the annihilation and creation rates is
equal to 2.  We can redefine the variable $\e$:
\[
\e= 1+ \frac{2}{\biggl(\frac{\hat{a}}{2-\hat{a}}\biggr)^y -1}
\]
where $\;y=1/2\;$.
The concentration for the finite chain is:
\ba
c(t,p,L) & = & \frac{1-\e}{2}+
\frac{(\e-\e^{-1})[1-(1-2p)^L\e^{-L}]}{2(1-\e^{-2L})} \nn\\[2mm]
& &          -\;\frac{1}{2L}\;\sum_{k=1}^{L-1}\bigg\{\biggl[
    \frac{1+(-1)^{k+1}(1-2p)^L}{\frac{1+(1-2p)^{2}}{2(1-2p)}-
\cos{\frac{\pi k}{L}}}-\frac{1+(-1)^{k+1}(1-2p)^L}{\frac{\e^2+1}{2\e}
-\cos{\frac{\pi k}{L}}}\biggr]\nn\\[2mm]
& & \hspace{2cm} \times \, \sin^{2}{\biggl(\frac{\pi k}{L}\biggr)}\exp
{(-\L^0_kt)}\biggr\}\;.
\label{ancre}
\ea
where $\L^0_k$ is:
\be
\L_k^0
  = \frac{2\e}{\e^2+1}\biggl(2(\e+\e^{-1})\:-4\:\cos\frac{\pi k}{L}\biggr)\;.
\ee
The steady-state concentration is $\;c_s=(1-\e^{-1})/2\;<\;0.5$.

After a permutation of the two lines in (\ref{ancret}) a system with
interchanged annihilation and creation rates is obtained. The
redefinition of $\e$ in the exactly solved case is the same as the one
given above but with $y=-1/2$. Formula (\ref{ancre}) gives the vacancies
concentration if $p$ is again replaced with $1-p$.
The steady-state concentration is $\;c_s=(1+\e^{-1})/2\;>\;0.5$.

In the limit $ \e \rightarrow \inf\;$ the reaction-diffusion system
in which all three rates are equal is obtained. The operator which
appears in the corresponding master equation is the Hamiltonian
of the Ising model \cite{Alc}.

\subsection {Three state models}
\label{3st}
\hspace{\parindent}
The starting point in this section is the two-species annihilation model.
The particles diffuse and
react when an $A$ and a $B$ try to occupy the same site:
\[
A\;+\;B\;\lra \;\nil\;+\;\nil.
\]
We denote the annihilation rate with $\,a\,$
($\,\G_{00}^{12}\,=\,a\,$), as in the preceding
section.
All diffusion rates are equal to 1 and $\,\G^{21}_{12}\,=\,0\,$ modeling
a hard-core interactions between particles.

The two sites Hamiltonian for this model is given by:
\vspace{0.5cm}
\be
H = 2 \times \le( \begin{array}{ccccccccc}
                0&0&0&0&0&-a&0&-a&0\\
                0&1&0&-1&0&0&0&0&0\\
                0&0&1&0&0&0&-1&0&0\\
                0&-1&0&1&0&0&0&0&0\\
                0&0&0&0&0&0&0&0&0\\
                0&0&0&0&0&a&0&0&0\\
                0&0&-1&0&0&0&1&0&0\\
                0&0&0&0&0&0&0&a&0\\
                0&0&0&0&0&0&0&0&0
    \end{array} \ri)\;
\label{ha3}
\ee
in the $\{|00\rangle,\;|01\rangle,\;|02\rangle,\,\cdots |22\rangle\}\,$ base.

A general transformation has the form (\ref{consu}):
\vspace{0.5cm}
\be
b = \le( \begin{array}{ccc}
         1-\a-\b & \g & \epsilon\\
         \a   & 1-\g-\d & \zeta \\
         \b   & \d & 1-\zeta-\epsilon
    \end{array} \ri)\;
\label{genf3}
\ee
We have to insure that the transformed Hamiltonian satisfies the
condition (\ref{sign}).
The first, fifth and ninth columns of $\hat{H}_i$
are identical up to a factor. So are the line one, five and nine.
The matrix elements of $\hat{H}$ can be interpreted in terms of reaction
rates only if the entries from at least one of these lines or from its
corresponding column are zero. The other two lines $i,j \in \{ 1,5,9 \}$
have at most two non-zero entries in the two corresponding columns
$i,j$. By imposing this conditions we are left with one transformation:
\vspace{0.5cm}
\be
b = \le( \begin{array}{ccc}
         1 & 1-\a &1-\b\\
         0   & \a & 0 \\
         0   & 0 & \b
    \end{array} \ri)\;
\label{part3}
\ee
which leads to a new reaction-diffusion system if $a\;$,$\a$ and $\b$ are
greater than one and if $\,\a^{-1}+\b^{-1} \geq 1\,$.
The transformed two sites Hamiltonian is:
\vspace{0.5cm}
\be
\hat{H} = 2 \times \le( \begin{array}{ccccccccc}
         0&0&0&0&0&-a(-1+\a^{-1}+\b^{-1})&0&-a(-1+\a^{-1}+\b^{-1}&0\\
         0&1&0&-1&0&-(1-\a^{-1})&0&-(a-1)(1-\a^{-1})&0\\
         0&0&1&0&0&-(a-1)(1-\b^{-1})&-1&-(1-\b^{-1})&0\\
         0&-1&0&1&0&-(a-1)(1-\a^{-1})&0&-(1-\a^{-1})&0\\
         0&0&0&0&0&0&0&0&0\\
         0&0&0&0&0&a&0&0&0\\
         0&0&-1&0&0&-(1-\b^{-1})&1&-(a-1)(1-\b^{-1})&0\\
         0&0&0&0&0&0&0&a&0\\
         0&0&0&0&0&0&0&0&0
    \end{array} \ri)\;
\label{hat2}
\ee

$\hat{H}\;$ has the same diffusion part and is left-right
symmetric. All processes which reduce the number of particles start
from a $\,A\,B\,$ pair. Apart from annihilation we have
'two species coagulation' reactions:
\begin{itemize}
\item	$\;\;\;\;A + B \ra \nil + \nil\;\;\;\;\;$ with rate
$\;\;\;\;\hat{\G}^{12}_{00}\,=\,a\,(\a^{-1}+\b^{-1}-1)$
\item	$\;\;\;\;A + B \ra \nil + A\;\;\;\;\;$ with rate
$\;\;\;\;\hat{\G}^{12}_{01}\,=\,1-\a^{-1}$
\item	$\;\;\;\;A + B \ra A + \nil\;\;\;\;\;$ with rate
$\;\;\;\;\hat{\G}^{12}_{10}\,=(\,a-1)(1-\a^{-1})$
\item	$\;\;\;\;A + B \ra B + \nil\;\;\;\;\;$ with rate
$\;\;\;\;\hat{\G}^{12}_{20}\,=\,1-\b^{-1}$
\item	$\;\;\;\;A + B \ra \nil + B\;\;\;\;\;$ with rate
$\;\;\;\;\hat{\G}^{12}_{02}\,=(\,a-1)(1-\b^{-1})$
\end{itemize}
Note that the diagonal terms of the Hamiltonian are invariant under the
transformation ($\;\hat{\G}_{12}\,=\,\G_{12}\,=\,a\;$).
The coagulation rates depend on the type of the particle which survives
the process and they are $(a-1)$ times smaller if the particle which survives
is the
one which jumps. There are only two coagulation rates,
independent of the initial position of the
surviving particle if $\,a=\,2\,$. This case corresponds to
the physical situation in which at any $A\;B$ encounter at least
one particle leaves the system.

{}From (\ref{eq:exvalnew}) and (\ref{conab}) it is easy to see that the
concentration will have an algebraic fall-off if:
\[
\frac{c_A(0)}{c_B(0)}\;=\;\frac{\a}{\b}.
\]

In this case the
particle concentrations are:
\be
c_A(t)\,=\a^{\frac 1 2} \frac{K\,\sqrt{c_A(0)}}{t^{1/4}}
\label{conan}
\ee
\hspace{0.5cm}
\be
c_B(t)\,=\b^{ \frac 1 2}\frac{K\,\sqrt{c_B(0)}}{t^{1/4}}\,=\,\frac{\b}{\a}
c_A(t)
\label{conbn}
\ee

\section {Monte Carlo Simulations}
\label{sec:mc}
\hspace{\parindent}
Only a few chemical models have been solved exactly. For some
others, quantitative estimates have been made with the help of
approximation schemes. This is the reason why numerical methods have been
extensively used in the study of reaction-diffusion systems.
The dynamics of the particle density and spatial distribution can be determined
with the help of Monte Carlo simulations \cite {ben-A, Doe}.

The Hamiltonian formalism opens new possibilities in this field.
For small chains numerical data can be obtained with a high accuracy
by using diagonalization techniques
or simulations. The study of the finite-size scaling behavior
(\ref{fssl}) in the limit $\, z \ra \nil\,$ permits the determination of
the particle concentration for infinite chains at very large times
\cite{Pap2}.

In this section we present our numerical results concerning the asymptotic
behavior of the particle concentration in the one and two species annihilation
models. We simulate directly the thermodynamical limit by using large lattices.

We consider chemical systems in which no reaction is permitted which creates
particles on a pair of empty sites. This is why the so called 'direct' Monte
Carlo Method \cite{ben-A, Arg} can be used. The way we implement this method is
described in detail in \cite{Pap2}.

We will first present our results
for the two state models and continue with the three state models.
\subsection{The two state annihilation-coagulation model}
\hspace{\parindent}The influence of the reaction rate in annihilation and
coagulation models is a subject which has received a
lot of attention in recent years (see \cite{S0, S1} and references
therein). We mentioned in the introduction the conjectured expression
(\ref{eq:ctb}) for the concentration decay in the $\,\D'\,\neq\,\nil\,$
coagulation model.  This result is in agreement with the conclusions
of a numerical study we carried out in our previous work. Using
a finite-size scaling analysis we obtained the same value for the time
exponents in the leading ($-1/2$) and the next to leading term ($-1$)
and we found that the coefficient of the leading term is independent of
$\D'$ \cite{Pap2}.

As mentioned there, a better test of (\ref{eq:ctb}) (or equivalently of
(\ref{eq:ltb})) can be made through a direct simulation of the
thermodynamic limit. Results of such computations are presented in \cite{S1}.
They are, in some cases, in semi-quantitative agreement with the
conjecture presented there. We concentrate on the long time behavior
and use smaller lattices as the one mentioned in \cite{S1}. Thus
we are able to obtain very good statistic in a reasonable $CPU$ time.

We tested the conjecture (\ref{eq:ctb}) in two different ways.
Simulations of the annihilation model were performed on a lattice of
length $L=2000$ for times up to a value of $t_{max}=1.5 \times 10^4$.

In the first set of simulations two values for the initial concentrations,
$1.0$ and $0.1$, were used. A decrease of 3 respectively 2 orders of magnitude
of the concentration enables us to reach the asymptotic regime.
300,000 runs were performed in order to insure a relative error of the
particle concentration of less then $10^{-3}$. In Figure 1 the following
quantity (see (\ref{eq:ltb}) with $f=0$):
\be
R(t) =\,2\pi\, \biggl( c(t) - \sqrt{\frac{1}{8\pi t}} \biggr)\frac{1-\D'}{\D'}
t
\label{rt}
\ee
is represented for $\D'=0.875$ and $0.75$. We see that our data converges to
the value $R(t)=1$ for which the straight line is drawn.

In Figure 2 the same kind of data corresponding to
$\D'=-7$ is shown. The convergence is worse because the absolute value of
the correction term (\ref{eq:ltb}) is $8$ times smaller than for $\D'=0.875$ so
$R(t)$ approaches faster the order of magnitude of the numerical errors.

A second set of simulations of the annihilation model were performed
in which we averaged only over 20,000 runs.
We applied a $\chi ^2$ test in two steps to the data. A linear
combination of time powers was presumed:
\be
y\,=\,K_x t^{-x}\,+\,K_{x+1/2}t^{-x-1/2}\,+\,K_{x+1}t^{-x-1}
\label{fitrel}
\ee
The first fit was made for the concentration $y=c(t)$ taking
$x=1/2$. The coefficient of the leading term was determined with a confidence
level greater then $99 \%$.  The first three significant digits of $K_{1/2}$
are identical with those of $(8 \pi)^{-1/2} \simeq 0.1995$.

The next step is to determine the coefficient of the first correction.
The leading term is subtracted from the numerical data:
\[
y\,=\,c(t)\,-\frac{1}{\sqrt{8 \pi t}}
\]
and a value $x=1$ is used in the  fit (\ref{fitrel}). The confidence level was
good (greater then $50\%$). The values for $R$ are listed in table
\ref{tab:2sr}.

\begin{table}[tb]
\centering
\begin{tabular}{|c|c||c|} \hline
\multicolumn{1}{|c|}{$\D'\,$} &
\multicolumn{1}{c||}{Initial concen-} &
\multicolumn{1}{c|}{Values of $R$} \\
\multicolumn{1}{|c|}{} &
\multicolumn{1}{c||}{tration $c(0)$} &
\multicolumn{1}{c|}{from M.C. data} \\ \hline \hline
$ -7    $&$ 1.00 $&$ 1.090 \pm 0.020$\\
$ -7    $&$ 0.10 $&$ 1.060 \pm 0.010$\\
$ -3/2  $&$ 1.00 $&$ 0.940 \pm 0.010$\\
$ -27/28$&$ 1.00 $&$ 0.950 \pm 0.020$\\
$ -1    $&$ 1.00 $&$ 1.000 \pm 0.040$\\
$ -1    $&$ 0.25 $&$ 1.180 \pm 0.080$\\
$ -1    $&$ 0.10 $&$ 1.060 \pm 0.080$\\
$ -3/7  $&$ 1.00 $&$ 1.010 \pm 0.030$\\
$ -1/9  $&$ 1.00 $&$ 0.940 \pm 0.070$\\
$  1/2  $&$ 1.00 $&$ 1.060 \pm 0.040$\\
$  1/2  $&$ 0.25 $&$ 0.990 \pm 0.040$\\
$  3/4  $&$ 1.00 $&$ 1.004 \pm 0.005$\\
$  3/4  $&$ 0.10 $&$ 1.030 \pm 0.020$\\
$  7/8  $&$ 1.00 $&$ 0.998 \pm 0.005$\\
$  7/8  $&$ 0.10 $&$ 1.005 \pm 0.005$\\
\hline
\end{tabular}
\protect
\caption{Values of $R$ (\protect\ref{eq:ctb}) obtained from Monte Carlo
simulations
of the annihilation model. The two different conjectured
values are $1$ \protect\cite{S1} and $0.786$ \protect\cite{S0} }
\label{tab:2sr}
\end{table}

For most values of $\D'$ simulations of the same model with different
initial concentrations were performed, as can be seen in table
\ref{tab:2sr}. The best fits were obtained for
the choice of reaction rates corresponding to a value of $\D'$ near
$1$. This is easy to understand because for such $\D'$s the contribution
of the next to leading term in (\ref{eq:ltb}) is maximal. The
simulations corresponding to $\D'\,=\,-7,\;3/4,\;$ and $7/8$  are the ones
used to make the Figures $1$ and $2$. For this data  the statistic was
made by averaging over 300,000 runs.

The conclusion we draw from the results of the two tests is that the conjecture
presented in \cite{S1}, predicting a value of $R=1$ in (\ref{eq:ctb}) is
correct.

\paragraph{} This results are valid if the data we considered before
are not affected by the finite size of the lattice on which the
simulations were performed. We check this by performing the simulations
described in the following.

In the asymptotic regime the system contains a low number of particles
which are spatially separated. A particle which was not involved in a
reaction represents a random walker. This means that at large times, of
the order of the square of lattice length, finite-size effects should be
dominant.

We obtain a more precise quantitative picture from simulations of
the annihilation model for times up to $\,t_{max}=10^6\,$.
We use a lattice of the same length ($\,L=2,000\,$) and averaged over 20,000
runs.  Three values for the parameter $\D'$ were chosen:$\,7/8\,$,$\,0\,$
and$\,-7\,$. Each simulation started with a full lattice as initial
configuration.

For an infinite system the long time behavior of the concentration is
algebraic .

For a finite system the concentration decay is exponential. We are
interested to study the transition between the two types of
decay, i.e. the onset of finite size effects at large times.

Figure 3 gives a log-log plot of the concentration. The straight line
corresponds to $\,c(t)\,=\,(8 \pi t)^{-1/2}\,$. We see that for times
$t\,>\,10^5\,$ the dashed lines strongly deviate from the straight line.

In Figure 4 which is a plot of the logarithm of the concentration we can see
that starting from the same value of time ($t\,=\,10^5\,$) the
curves fit nicely with another straight line described
by:
\be
c(t)\,=\,\frac 2 L \exp{\biggl(-\,2\,\frac{\pi^2}{L^2}t\biggr)}\;.
\label{fse}
\ee
The formula is obtained in the $\,t\rightarrow\inf\,$ limit of
(\ref{eq:genkonz}), for the
$\,\D'\,=\,0\,$
annihilation model $(\,c\,=\,0\,)$.

{}From Figures 3 and 4 we conclude that the onset
of the finite-size effects takes place in a narrow vicinity
of $t\,=\,10^5\,$. This  value is considerably smaller than $L^2$ but much
larger than the $t_{max}$ we used. It is now clear that the results
we presented in the first part of this section refer to the thermodynamic
limit of the annihilation model.

The slope of the line from Figure 4 is equal to the energy of the first excited
state of $H_{\mbox{\scriptsize ann.}}$. We can conclude that this energy has
the
value $\,4(1-\mbox{cos}(\pi/L))\,$ independent of the parameter $\D'$.
This confirms previous results obtained from numerical diagonalization
of this Hamiltonian \cite{Malte}.
\subsection{Numerical results for the three state models}
\hspace{\parindent}

The simulation of the thermodynamic limit of the two species
annihilation reaction
requires a very large amount of {\em CPU} time.

This has two reasons. A reliable study of the
long time decay of the concentration presumes simulations
in which the total number of particles reduces by at least two orders of
magnitude. The algebraic fall-off is much slower than in the $\;A+A \lra
\nil+\nil\;$ case. This means that we have to perform the simulations
up to much higher values of $t_{max}$ than we did in the one species case.

On the other hand, if we start at $\,t\,=\,0\,$ with a random
distribution, there will be local fluctuations of the particle densities.
The decay of the initial fluctuations  plays an essential role in the dynamics
of the system
\cite{T1}. Their length scale extends in time. So
one is forced to use longer lattices than in the one-species case where
such domains do not appear.

A lattice of length $\,L\,=\,10^5\,$ was used to simulate three state
models. In the case of the two species annihilation we were able to
average only over $100$ runs.
Our aim is to see if the particle concentration obeys the algebraic
decay law (\ref{conab}) for the $A-B$ symmetric case and to compute
$K$. Results from similar simulations were reported in
\cite{T1} which lead to $K=0.280$. This is in equally
good agreement with the two conjectured values $0.252$ and $0.316$.

Figure 5 shows the curve of
\be
K(t)\,=\,c_A(t)\,\frac {t^{1/4}}{\sqrt{c(0)}}
\label{kt}
\ee
for various choices of the annihilation rate and initial site occupation
probabilities. The simulations were stopped at $t_{max}=10^6$, a value hundred
times larger as the one used in \cite{T1}. At very large times the curves
converge showing that the asymptotic regime was reached. The average
value is $K=0.247 \pm 0.004\,$.

Our result is important because it confirms the conjectured dependence
of the concentration on its initial value \cite{T1}. It also suggest that the
asymptotic behavior is universal in the sense that it is independent of the
annihilation rate $a$.

We cannot make any prediction concerning the exponent $x$ in (\ref{conab}).
This would require a better statistic
in the determination of the asymptotic behavior of the concentration.
This is an aim which we cannot achieve with the computer facilities
which are currently at
our disposal.

\section* {Conclusions}
\hspace{\parindent} We found a general condition satisfied by local
transformations which realize mappings between reaction-diffusion systems.
They are $N \times (N-1)$-parameter real matrices which have the sum of
entries in each column equal to one.

Some new results concerning two state models are derived:

All annihilation-coagulation systems corresponding to the same
parameter $\,\D'\,$ (\ref{d'}) are equivalent. The continuous similarity
transformation which acts between these models is given. Results
previously derived for the coagulation and annihilation models are generalized
to the chemical systems in which both reactions are permitted.

At this point we return to the exciton decay measurements mentioned in
the introduction. Further experimental determinations are very desirable
for two reasons:

1) It would be interesting to have accurate measurements of the
exciton hopping time which is related to their diffusion
constant. The coefficient of the leading term ($\; \sim t^{-1/2}\;$) in
the exciton concentration decay can then be extracted from the experimental
data \cite{S11}. Thus one can determinate the ratio of the exciton-exciton
annihilation and coagulation rates(\ref{eq:ltb}).

2)Lower errors in the experimental measurements of the luminescence decay
would permit the computation of the next to leading term (its time exponent
and coefficient) which appears in an expansion of the exciton concentration.

If this term is of order $\;t^{-3/2}\;$ (\ref{eq:slt}) the dynamic of
exciton propagation and reactions is described by a system which is
supersymmetric \cite{Alc}. This would be the first such system
experimentally measured. The other possibility is that the next to leading
term is of order $\;t^{-1}\;$ (\ref{eq:ltb}). In both cases
the determination of the leading and next to leading term and of the hopping
time would permit the computation of the rates of the processes which
take place between excitons.

The two states coagulation-decoagulation, annihilation-creation and
death-decoagulation models are equivalent. This enables us to give
an exact expression for the particle number of a probabilistic cellular
automaton in which modifications of the configuration occur only by starting
from particle-vacancy pairs (\ref{eq:coagkonz}). The same observable is
computed for the model in which the sum of the annihilation and creation
probabilities is one (\ref{ancre}).

A reaction-diffusion system equivalent to the two species annihilation
($\; A+B \ra \nil + \nil \;$) is found.

\hspace{0.5cm}

We have accurately determined the effect of the reaction
rate on the long time decay of the concentration in the one-species
annihilation model. Our results are in excellent agreement with a theoretical
conjecture previously made in \cite{S1}.

We also present quantitative analysis of the asymptotic behavior
of concentration in the three states annihilation model.
They are also fully compatible with previous results. Although they
are, as far as we  know, the best available in the literature, they
do not allow us to make any prediction concerning the effects of the
reaction rate as we did for the $A+A \lra \nil + \nil $ model.

\section* {Acknowledgements}
\hspace{\parindent} I am grateful to Prof. V. Rittenberg for very helpful
discussions during the course of this work. I would like to thank P.
Arndt, T. Heinzel, K. Krebs and B. Wehefritz for useful discussions and
for carefully reading the manuscript.

\newcounter{fig_count}
\vspace{1cm}
\noindent {\Large \bf List of Figures}

\vspace{1cm}

The Monte Carlo data are represented by interrupted lines. Error bars are
given only for a few points.

\begin{list}
      {Fig. \arabic{fig_count}:}
      {\usecounter{fig_count}  \setlength{\leftmargin}{1.5cm}
               \setlength{\labelsep}{2mm}
               \setlength{\labelwidth}{1.3cm}}
\item Effects of the annihilation rate on the asymptotic behavior
of the particle concentration for the $A+A\lra \nil + \nil$ reaction
$\;\;-\;\;R(t)$ as defined by (\ref{rt}), for
$\D'\,=\,$ $3/4$ and $7/8$. The straight line is $R(t)\,=\, 1$.
\item Effects of the annihilation rate $(2)$.
$R(t)$ as defined by (\ref{rt}), for $\D'\,=\,$ $-7$.
\item Onset of finite size effects - the algebraic decay.
Double logarithmical plot of the concentration for the  one-species
annihilation reaction.
\item Onset of finite size effects - the exponential decay.
Plot of the logarithm of the concentration for the one-species
annihilation reaction.
\item Asymptotical behaviour of the particle concentration in the
$A+B \lra \nil + \nil$ reaction. Plot of $K(t)$ as defined
by (\ref{kt}). The straight line corresponds to $K(t)=(2 \pi)^{-3/4}$
\end{list}
%

\begin{thebibliography}{99}
\bibitem {bA2}     D. ben-Avraham, M.A. Burschka and C.R. Doering,
                   {\em J. Stat. Phys.} {\bf 60}:695 (1990).
\bibitem{Lush}     A.A. Lushnikov,
                   {\em Sov. Phys. JEPT } {\bf 64}:811 (1986).
\bibitem{Alc}      F.C. Alcaraz, M. Droz, M. Henkel and V. Rittenberg,
                   {\em Ann. Phys.} {\bf 230}:250 (1994).
\bibitem{Pap1}     K. Krebs, M.P. Pfannm\"uller,
                   B. Wehefritz and H. Hinrichsen,
                   {\em J. Stat. Phys.} {\bf 78}:1429 (1995)
\bibitem{Kadanoff} L.P. Kadanoff and J. Swift,
                   {\em Phys. Rev.} {\bf 165}:165 (1968).
\bibitem{S8}       R. Kopelman,
                    {\em J. Stat. Phys.} {\bf 42}:185 (1986).
\bibitem{S9}       R. Kopelman, S.J. Parus and J. Prasad,
                    {\em J. Chem. Phys.} {\bf 128}:209 (1988).
\bibitem{S10}      R. Kroon, H. Fleurent and R. Sprik,
                     {\em Phys. Rev.} {\bf E} {\bf 47}:2462 (1993).
\bibitem{S11}      W.J. Rodriguez, M.F. Herman, G.L. McPherson,
                        {\em Phys. Rev.} {\bf B} {\bf 39}:18 (1989).
\bibitem{S0}       D. Zhong and D. ben-Avraham,
                   {\sl Diffusion-Limited Coalescence with Finite
                   Reaction Rates in One Dimension},
                   {\em preprint} (1994).
\bibitem{S1}       V. Privman, C.R. Doering and H.L. Frisch,
                   {\em Phys. Rev.} {\bf E} {\bf 48}:846 (1993).
\bibitem{Privman}  V. Privman,
		   {\em Phys. Rev.} {\bf E} {\bf 50}:50 (1994).
\bibitem{Kang}     K. Kang and S. Redner,
                   {\em Phys. Rev.} {\bf A} {\bf 30}:2833 (1984).
\bibitem{Spouge}   J. Spouge,
                   {\em Phys. Rev. Lett.} {\bf 60}:871 (1988).
\bibitem{Kang2}     K. Kang and S. Redner,
                   {\em Phys. Rev. Lett.} {\bf 52}:955 (1984).
\bibitem{T1}       D. Toussaint, F. Wilczek,
                   {\sl Particle- antiparticle annihilation in diffusive
                      motion},
                   {\em J. Chem. Phys.} {\bf 78}:2642 (1983).
\bibitem{T3}       B.P. Lee,
                   {\sl Critical Behaviour in Non-Equilibrium Systems},
                   {\em Ph. D. Thesis}, University of California, Santa
                        Barbara, (1993).
\bibitem{T2}       J.-C. Lin, {\em Phys. Rev.} {\bf A} {\bf 44}:6706
                   (1991).
\bibitem{PRS}      I. Peschel, V. Rittenberg, U. Schultze,
                   {\em Nucl. Phys.} {\bf B} {\bf 430}:633 (1994).
\bibitem{Doering62}  C.R. Doering and D. ben-Avraham,
                     {\em Phys. Rev. Lett.} {\bf 62}:2563 (1989).
\bibitem {DBH}     C.R. Doering, M.A. Burschka and W. Horsthemke,
                   {\em J. Stat. Phys.} {\bf 65}:953 (1991).
\bibitem{Derr}     E. Derrida,
                   {\sl Exponents Appearing in the Zero Temperature
                        Dynamics of the 1d Potts Model},
                   {\em preprint} 1994.
\bibitem{ben-A}     D. ben-Avraham,
                   {\em J. Chem. Phys.} {\bf 88}:941 (1988).
\bibitem{Doe}  C.R. Doering and D. ben-Avraham,
                   {\em Phys. Rev.} {\bf A} {\bf 38}:3035 (1988).
\bibitem{Pap2}     K. Krebs, M. Pfannm\"uller, H. Simon and
                   B. Wehefritz,
                   {\em J. Stat. Phys.} {\bf 78}:1471 (1995)
\bibitem{Arg}       P. Argyrakis,
                   {\em Computers in Physics} {\bf 6}:525 (1992).
\bibitem{Malte}    M. Henkel, {\em unpublished}
\end{thebibliography}
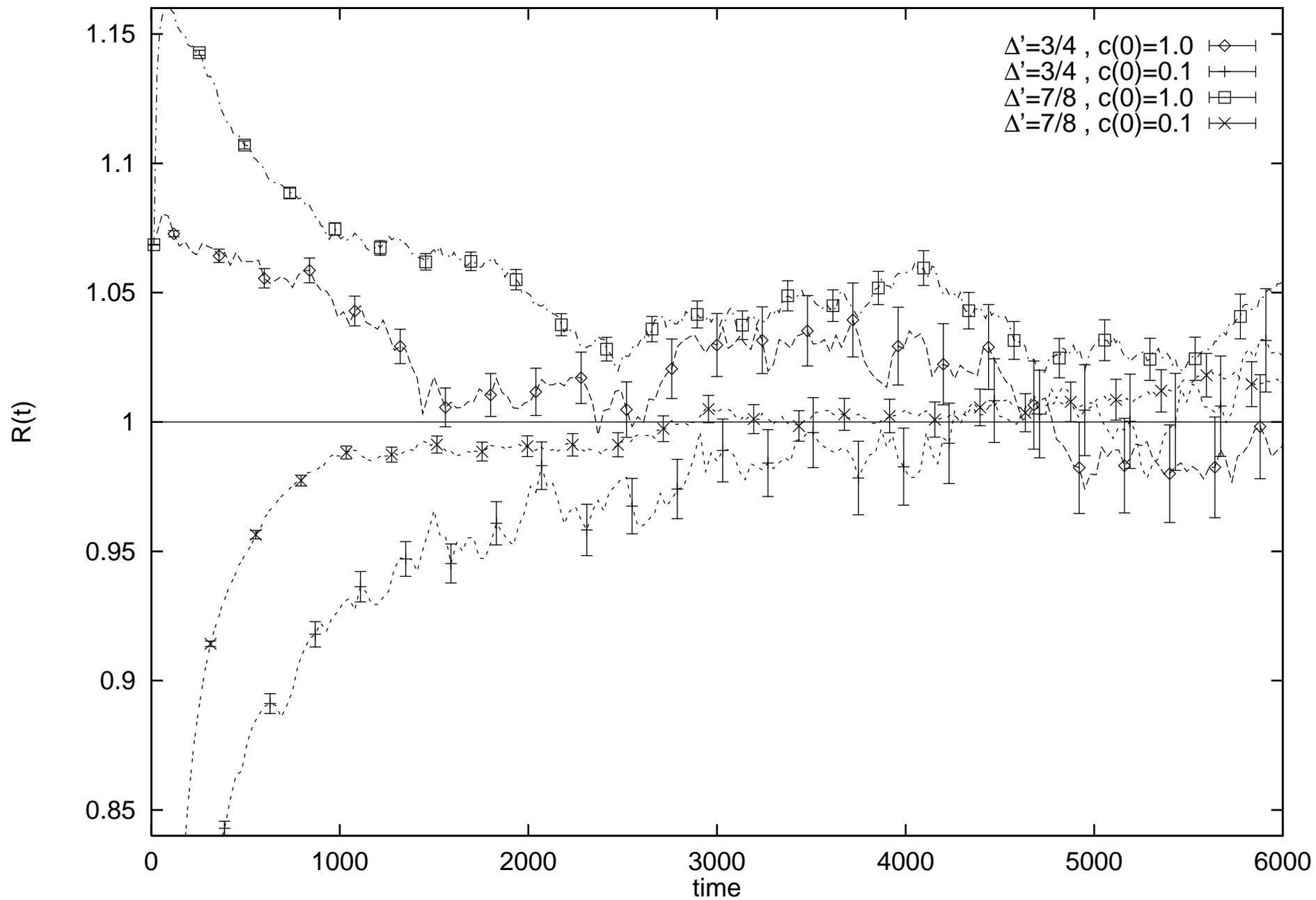
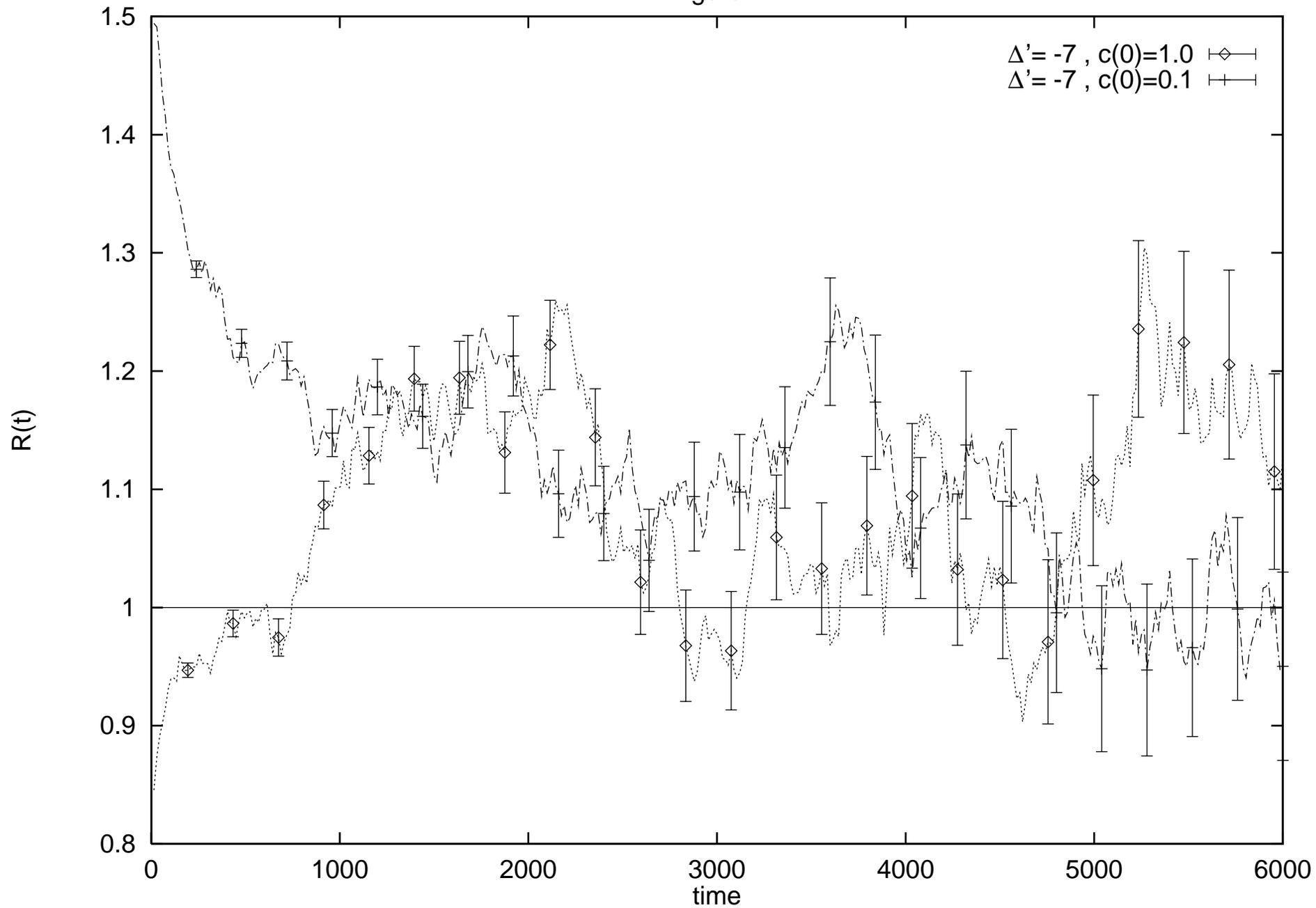
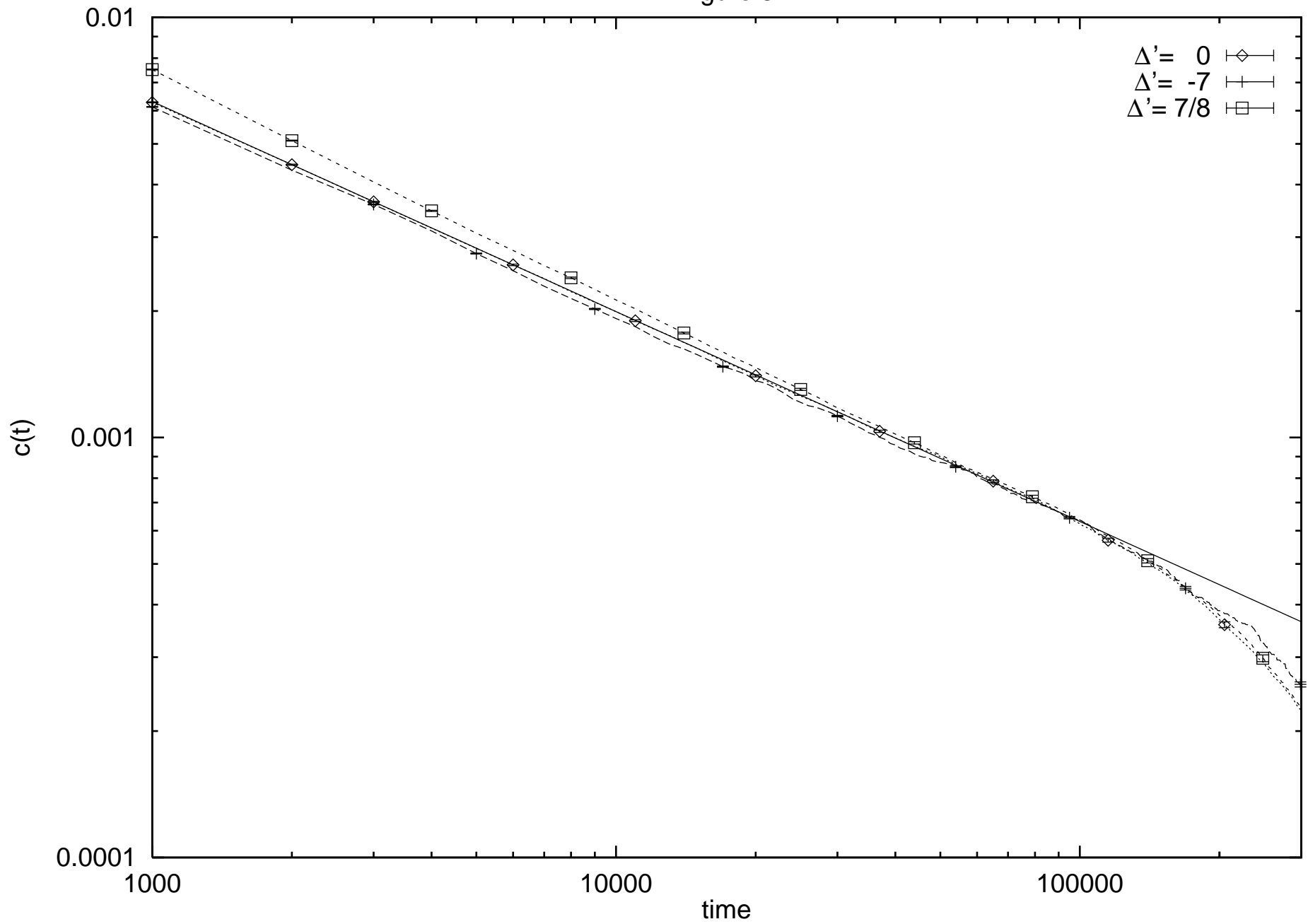
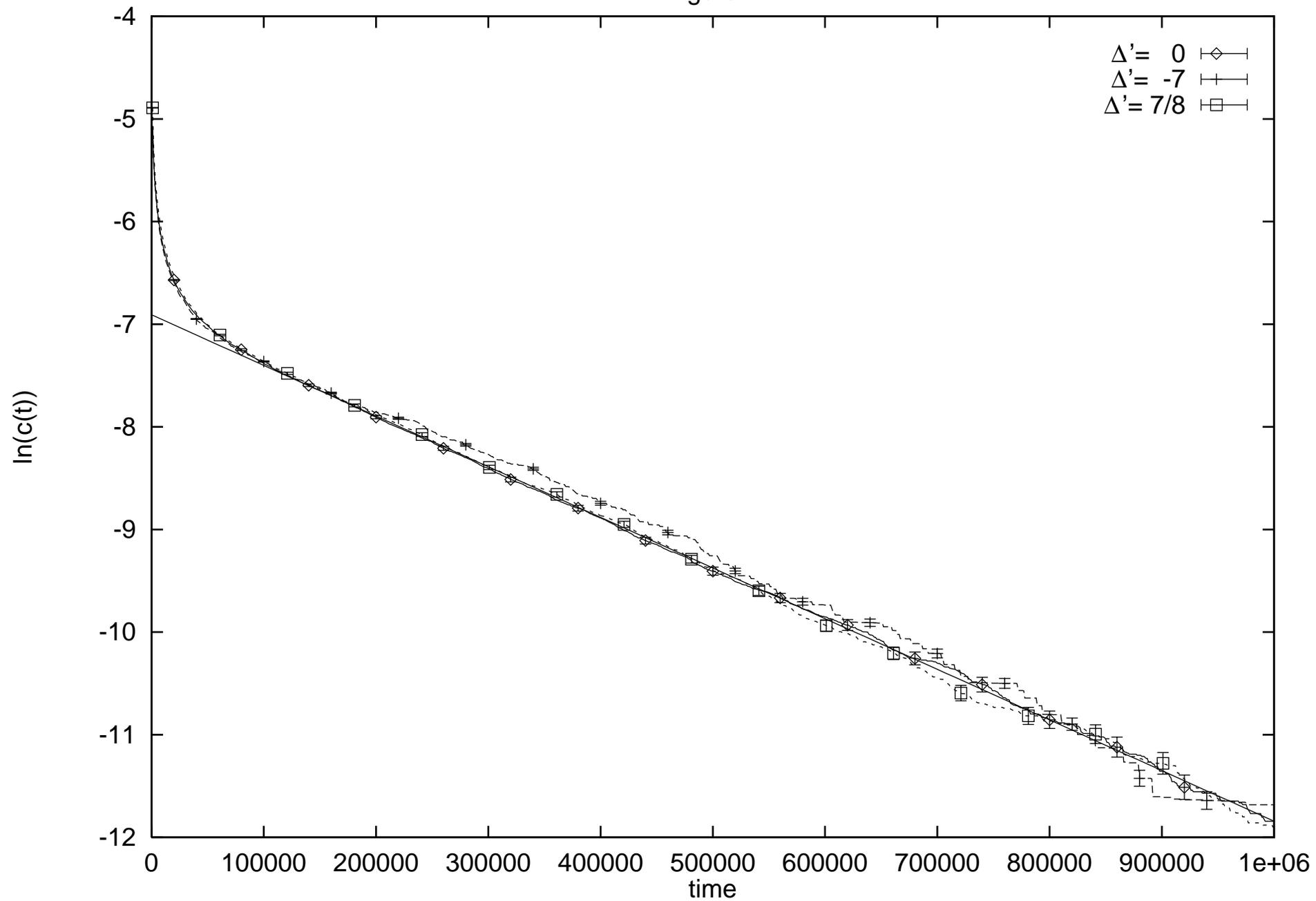
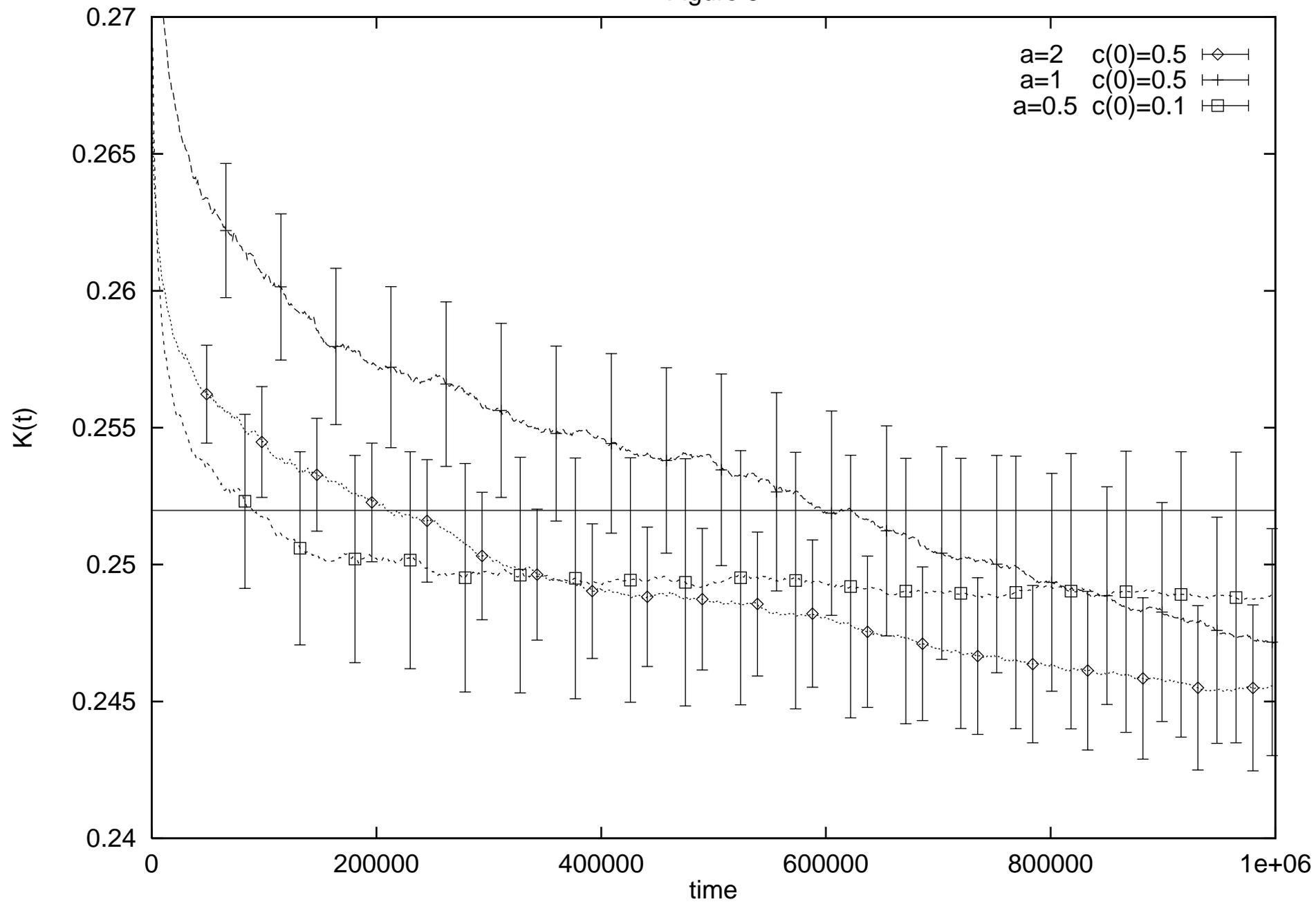
\end{document}